\documentclass[aps,prl,twocolumn,superscriptaddress]{revtex4}
\usepackage{bm}
\usepackage{epsfig}
\usepackage[english]{babel}
\usepackage{graphicx,times}
\usepackage{amssymb,amsmath,bbm,amsfonts,amsthm,subfigure,dsfont,palatino,txfonts}

\newcommand{\abs}[1]{\left\vert#1\right\vert}

\newcommand{\ket}[1]{\left\vert#1\right\rangle}

\begin{document}

\title {The driven Dicke Model: time-dependent mean field and quantum fluctuations in a non-equilibrium quantum many-body system}
\author{G.~Francica}
\affiliation{Dip.  Fisica, Universit\`a della Calabria, 87036
Arcavacata di Rende (CS), Italy} \affiliation{INFN - Gruppo
collegato di Cosenza, Cosenza Italy}
\author{S.~Montangero}
\affiliation{Institute for Complex Quantum Systems, Ulm University, Albert-Einstein-Allee 11, 89069 Ulm, Germany}
\author{M.~Paternostro}
\affiliation{Centre for Theoretical Atomic, Molecular and Optical
Physics, School of Mathematics and Physics, Queen's University,
Belfast BT7 1NN, United Kingdom}
\author{F.~Plastina}
\affiliation{Dip.  Fisica, Universit\`a della Calabria, 87036
Arcavacata di Rende (CS), Italy} \affiliation{INFN - Gruppo
collegato di Cosenza, Cosenza Italy}

\begin{abstract}
We establish a new theoretical framework, based on a
time-dependent mean field approach, to address the dynamics of the
driven Dicke model. The joint evolution of both mean fields and
quantum fluctuations gives rise to a rich and generally non-linear
dynamics, featuring a normal (stable) regime and an unstable,
super-radiant one. Various dynamical phenomena emerge, such as the
spontaneous amplification of vacuum fluctuations, or the
appearance of special points around which the mean-field
amplitudes rotate during driven time evolution, signalling a
dynamical symmetry breaking. We also provide a characterization of
the driving-induced photon production in terms of the work done by
the driving agent, of the non-adiabaticity of the process and of
the entanglement generated between the atomic system and the
cavity mode.
\end{abstract}
\maketitle The dynamical behavior of quantum critical systems
displays interesting features concerning defects or excitations
production~\cite{polkormp,eisertnatphys}, which occurs when the
system is driven across its critical point \cite{varikz}. The
Dicke model is a paradigmatic example in this context, embodying a
quantum many-body system with highly non-trivial critical
features~\cite{Brennecke,pnas}, where an electromagnetic mode is
coupled to a collection of $N$ identical two-levels
atoms~\cite{dicke}. Indeed, strong correlations are set both among
atoms and with the field, which in turn result in significant
non-classical behavior of the radiation and in cooperative effects
giving rise to both atomic and photon squeezing.

These tantalising features persist in the case of external driving
of the dynamics, as remarkably shown experimentally in
Ref.~\cite{Brennecke}, where the corresponding spontaneous
symmetry-breaking effect induced by an adiabatic crossing of the
quantum critical point of the model  has been demonstrated.
Non-adiabatic driving has also been the focus of substantive
theoretical and experimental investigation~\cite{varie}. In
particular, for a periodic driving of the atom--field coupling
strength, Ref.~\cite{Bastidas12PRL108} has shown the emergence of
new metastable phases in the driven Dicke model, whose phase
diagram appears to be substantially different from the static one,
both qualitatively and quantitatively.

In this paper we discuss symmetry breaking and photon generation
in the driven Dicke model, where the frequency of the field, the
energy of the atoms, {\it and} their mutual coupling are all
allowed to vary in time. While the dynamical behavior of the
single-atom case has been recently studied~\cite{plenion=1}, we
are interested in the thermodynamic limit. By non-adiabatically
driving the system, and thoroughly analysing its stability
conditions, we highlight implications that non-adiabaticity has on
the evolution of both the mean fields and the residual quantum
fluctuations. Indeed, the {\it macroscopic} trajectories followed
by the mean fields are shown to dynamically select one of the
symmetry broken configurations each time the critical point is
crossed; while the {\it microscopic} features of the fluctuations
are shown to be crucial for the characterization of the temporal
behavior of key observables of the system, such as the photon
number. Moreover, we study the thermodynamic work produced by
driving the system, and the associated degree of irreversibility,
thus addressing the non-adiabatic production of photons from the
electromagnetic vacuum, a phenomenon akin to the dynamical Casimir
effect~\cite{Vacanti12PRL108}, and the associated generation of
atom-field entanglement, from a genuinely non-equilibrium
perspective.

For a negligible mutual atomic interaction and for an atomic
system occupying a linear dimension much smaller than the
electromagnetic wavelength, the atom field system is described by
the collective-spin Hamiltonian $\hat H= \hat H_0 + \hat H_{int}$
with
\begin{equation}\label{hami}
\hat H_0 = \omega_a \hat a^\dag \hat a + \omega_b \hat J_z \, , \quad \hat H_{int}= {2
g} \left( \hat a^\dag + \hat a \right) \hat J_x/\sqrt N.
\end{equation}
Here $\hat a$ ($\hat a^\dag$) is a bosonic annihilation (creation) operator, and  $\hat {\bm J} = (
\hat J_x, \hat J_y, \hat J_z  )$ is the collective spin operator, with $\hat{\bm J}
= \sum_{i=1}^N {\hat{\bm \sigma_i}}/{2}$ and $\hat{\bm \sigma_i}$ is the vector of Pauli spin operators.

As mentioned, we will be interested in the case where
$\omega_{a,b}$, and $g$ are all functions of time (in order to
avoid notational clutter, and unless otherwise specified, we will
avoid writing explicitly any time dependence). In fact, as we
discuss below, the key ingredient in the dynamics of the system is
the time dependence of the parameter $\mu={\omega_a \omega_b}/(4
g^2)$. We will treat the atoms as indistinguishable, and consider
a fully symmetric initial atomic state. This feature is preserved
during time evolution as the symmetric subspace is dynamically
invariant. We now consider the Holstein-Primakoff transformation
of $\hat{\bm J}$ restricted to such subspace and thus introduce
the bosonic operators $\hat b$ and $\hat b^{\dag}$ such that
\begin{equation}
\label{HP}
\hat J_z =\hat b^\dag \hat b - N/2  \, , \quad
  \hat J_+ \equiv \hat J_x + i \hat J_y =\hat b^\dag \sqrt{N - \hat b^\dag \hat b}.
\end{equation}
As a result, within the symmetric subspace $\hat H_{int}$ takes
the form
\begin{equation}\label{hamiltonian HN}
 \hat H_{int}=
g \left(\hat a^\dag +\hat a \right)\left(\hat b^\dag\sqrt{1-\frac{\hat b^\dag \hat b}{N}}  + \sqrt{1-\frac{\hat b^\dag \hat b}{N}}\hat b \right).
\end{equation}
In the time-independent case and at the thermodynamic limit, $\hat
H$ can be diagonalized by isolating from $\hat b$ a macroscopic
($\sim\sqrt N$) mean contribution~\cite{Emary03PRE67} and
retaining only the leading terms of a $1/N$ expansion of the
square roots in Eq.~\eqref{HP}, thus obtaining a quadratic
Hamiltonian. This procedure implies subtracting a {\it static}
mean field chosen in order to approximate the Hamiltonian as
accurately as possible at low energies, or, loosely speaking,
chosen in such a way as to minimize residual quantum fluctuations
near the ground state. Our approach to the investigation of the
driven model is based on an analogous idea: we will isolate {\it
time-dependent} mean fields, chosen so as to minimize residual
quantum fluctuations around the {\it instantaneous} state vector
$\ket{\psi,t}$, whose evolution is generated by an approximately
quadratic time-dependent Hamiltonian.

In the thermodynamic limit and for $\mu > \mu_c=1$ [$\mu<1$],
$\hat H$ admits a normal (N) [super-radiant (SR)] quantum phase.
At $\mu=1$, a second-order phase transition is found. As we
discuss below, the driven system correspondingly displays two
dynamical regimes~\cite{Bastidas12PRL108}.

In order to perform a quantitative analysis, we start by shifting
the field and atomic operators $\hat a$ and $\hat b$ by their
time-dependent mean values $\langle \hat a \rangle = \sqrt{N}
\alpha$ and $\langle \hat b \rangle = \sqrt{N} \beta$, thus
introducing new operators describing deviations from the averages,
$\hat c = \hat a - \sqrt N \alpha$, and $\hat d = \hat b - \sqrt N
\beta$ (with $\alpha,\beta\in\mathbb{C}$). We require any
macroscopic contribution to be ascribed to the mean fields, and
thus assume that fluctuations remain very small. Therefore, after
the rescaling by $\sqrt{N}$ performed above, we expect  $\alpha$
and $\beta$ to remain ${\cal O}(1)$ as $N\rightarrow \infty$ (as
it is the case for a static $\hat H$~\cite{Emary03PRE67}), see
\cite{nota} for details.
We thus have 
\begin{equation}
\label{expansion}
\sqrt{1-\frac{\hat b^\dag \hat b}{N}}
\simeq \sqrt{\Gamma}\bigg[ 1 - \frac{\beta \hat d^\dag + \beta^* \hat d}{2 \Gamma \sqrt{N} } -\frac{d^\dag \hat d}{2N \Gamma} 
-\frac{(\beta \hat d^\dag + \beta^* \hat d)^2}{8 N
\Gamma^2}\bigg]
\end{equation}
with $\Gamma = 1-\abs{\beta}^2$. Using the leading terms only, one
can derive equations for the
bosonic operators and their averages 
as~\cite{nota}
\begin{equation}\label{eqs}
i  \dot{\alpha} =  \omega_a \alpha + 2 g \sqrt{\Gamma} \beta_{r},\,\, i \dot{\beta} =  \omega_b \beta + 2 g \sqrt{\Gamma}
\alpha_{r}\left(1-{\beta \beta_{r}}/{\Gamma}\right).
\end{equation}
Here $s_r=\text{Re}(s)$ and $s_i=\text{Im}(s)$ with
$s=\alpha,\beta$. Eq.~\eqref{eqs} are explicitly nonlinear.
However, by starting from low-energy conditions, the dynamics can
be well approximated by linear equations of motion up until the
time $t_{\text{lin}}$ within which $|\alpha|,|\beta|\ll
1$~\cite{nota}. When the mean fields acquire macroscopic values,
their dynamics become fully non linear and the whole of
Eq.~(\ref{eqs}) should be retained. For a constant Hamiltonian and
any value of $\mu$, these equations admit the stationary solution
$\alpha^n=\beta^n=0$, corresponding to the ground state of the N
phase~\cite{Emary03PRE67}. In the SR-phase, with $\mu <1$, two
other stationary points appear, corresponding to states with
broken (parity) symmetry,
$\alpha^{sr}_{\pm} = \pm  \frac{g}{\omega_a}\sqrt{1-\mu^2}$, and
$\beta^{sr}_{\pm} = \mp \sqrt{\frac{1-\mu}{2}}$.
For a driven system, $\mu$ depends on time and so do $\alpha^{sr}$
and $\beta^{sr}$, while the normal value remains stationary. As a
result, if the initial state has null mean fields, the condition
$\alpha=\beta=0$ will hold at all times and the dynamics never
exit the linear transient. However, if the system is
instantaneously brought into the SR region, such normal stationary
state can become unstable.

\begin{figure}[t]
        \begin{center}
        \includegraphics[width=\linewidth]{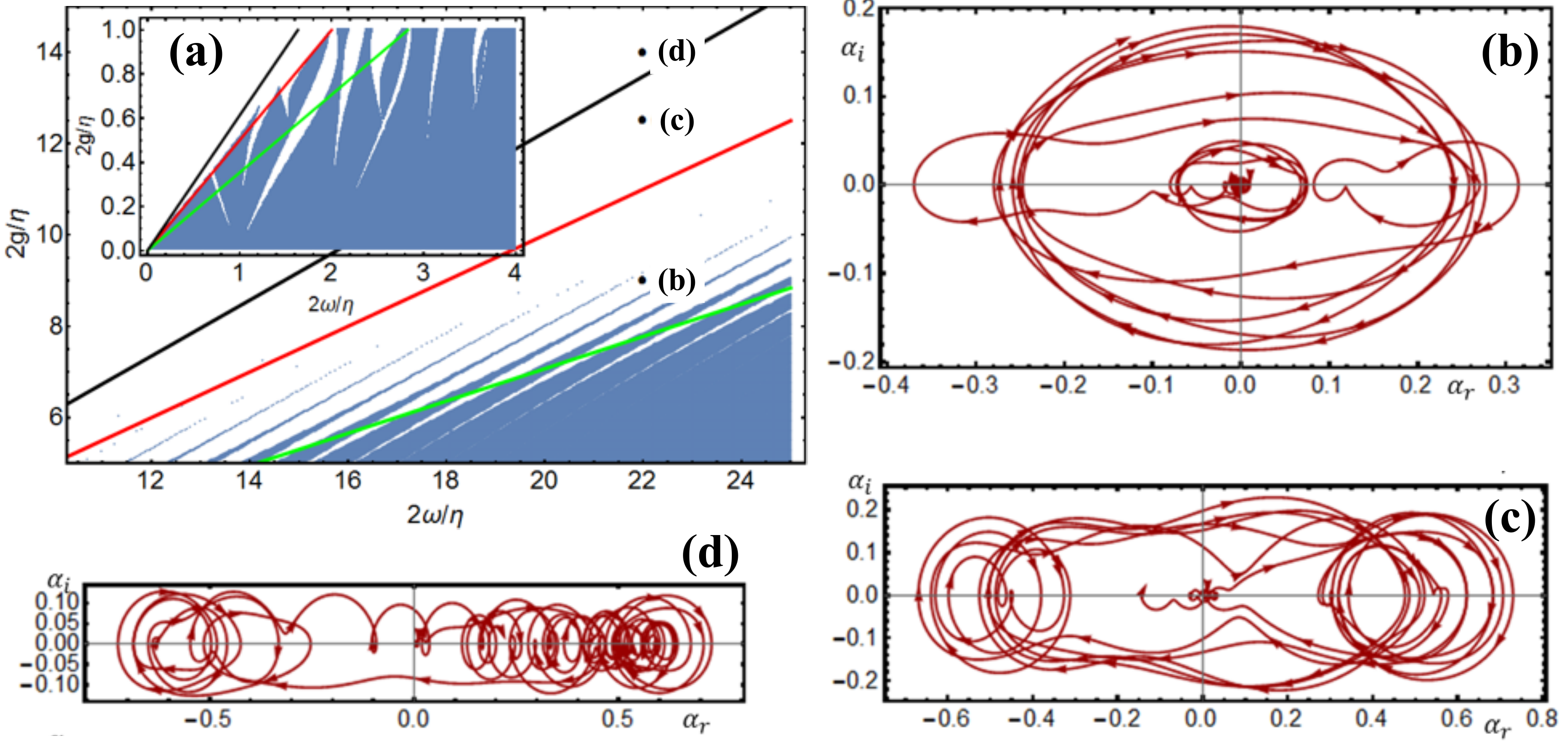}
\caption{(Color online) {\bf (a)} Stability diagram in the driving
frequency vs. coupling plane, signalling in white the regions with
positive instability rate, $\gamma^*>0$. The plot is drawn at the
static resonance, $\lambda_0=1$, with $\lambda=\frac{1}{2}$.
The red line marks the static critical coupling 
$\mu_0=1$, while the green one corresponds to
$\mu_{min}=1$; so that, for a point under the green line $\mu(t)>1
\forall t$, while $\mu(t)$ periodically goes below unity above it.
Finally, the black line corresponds to $\mu_{max}=1$. For
$g\rightarrow 0$ the white zones open for $\eta_k= 2 \omega_a /k$,
$k\in Z$. In the limit $\eta \rightarrow \infty$ the first blue
triangle on the left tends to fill the whole region under the red
line, while for $\eta \rightarrow 0$ it is the region under the
green line that becomes blue. The inset shows the stability
diagram for larger values of the driving frequency. {\bf (b)}-{\bf
(d)} Trajectories of the mean field $\alpha$ corresponding to the
three points in panel {\bf (a)}. We have assumed a slow periodic
driving ($\omega_a = 11 \eta$) and evaluated $\alpha(t)$ up to $t
= 60 \eta^{-1}$ by numerically solving Eqs.~(\ref{eqs}) for the
parameters indicated by the black dots in the central panel, with
$2 g /\eta = 9, 12.5, 14$. Initially, the radiation field has been
taken in the coherent state $|{\sqrt{N}\epsilon}\rangle$
with $\epsilon = 10^{-2}$, with all of the atoms in their ground states. 
}
        \label{Fig1}
        \end{center}
\end{figure}
In the following, we focus on a periodically driven system.
We specifically assume the atomic frequency to be sinusoidally
perturbed, $\omega_b/\omega_a = \lambda_0 + \lambda \sin(\eta t)$,
as in ~\cite{Vacanti12PRL108}, implying an harmonic time
dependence for $\mu(t)$, which oscillates with frequency $\eta$,
between $\mu_{min}=\mu_0 (1- \lambda/\lambda_0)$ and $\mu_{max}=
\mu_0 (1+ \lambda/\lambda_0)$, where $\mu_0=\lambda_0 \omega_a^2/4
g^2$. For such a periodic driving, we can make use of Floquet
theory~\cite{Floq} to study the dynamics of the system. In
particular, the stability of the solutions of Eqs. (\ref{eqs}) can
be characterized by an instability rate $\gamma^*$ defined as the
largest positive Floquet exponent of the linearized
equations~\cite{Floq}, which embodies the growing rate of the mean
fields in the linear regime. While the details of this analysis
are given in the supplementary material~\cite{nota}, the result is
reported in Fig.~\ref{Fig1}, where we see that $\gamma^*>0$ if
$\mu(t)< 1$ at all times (above the black line in
Fig.~\ref{Fig1}), while driving-induced instabilities appear even
in the (static) N region (below the red line in Fig.~\ref{Fig1}).
For small couplings, this occurs near the parametric resonance
points $\eta_k = \omega_a(1+\lambda_0)/k$ ($k\in\mathbb{Z}$).
These are the so called Arnold instability tongues, discussed in
\cite{Bastidas12PRL108}. Although our treatment is valid in
general, the discussion below will mainly focus on the case of a
slow (but non-adiabatic) driving, therefore Fig.~\ref{Fig1}
explicitly displays the case $\eta \ll \omega_a, \omega_b$.

Although the dynamics can exit the linear regime when the mean
fields acquire large values (which can occur quite quickly, e.g.,
for initial coherent states with very small amplitudes), the
diagram in Fig. \ref{Fig1} still helps classifying the dynamical
behavior of the mean fields, identifying those values of frequency
and coupling for which $\alpha$ and $\beta$ grow exponentially in
time, from those for which they stay bounded. As shown below, the
very same diagram will help in the study of quantum fluctuations
(cfr. the discussion after Eq. \ref{wveceq}).

Solutions of Eqs.~(\ref{eqs}) in the different regimes are
reported in Fig.~\ref{Fig1} {\bf (b)}-{\bf (d)}, where various
examples of trajectories of the photon mean field $\alpha$ are
shown ($\beta(t)$ follows similar paths).
At $t=0$ the electromagnetic mode is taken in the coherent state
$|{\sqrt{N}\epsilon}\rangle$  (with $\epsilon \ll 1$), while the
atoms are in their ground states. If the system's parameters are
chosen in the unstable region, the mean fields grow exponentially
and, once out of the linear regime, get macroscopic values. For a
set of parameters in the stable zone, instead,  the trajectory is
bounded, with $|\alpha(t)|$ remaining $ \sim \epsilon$. This
behavior is quite general and does not depend on the value of the
driving frequency.

\begin{figure}[t]
        \begin{center}
        \includegraphics[width=\linewidth]{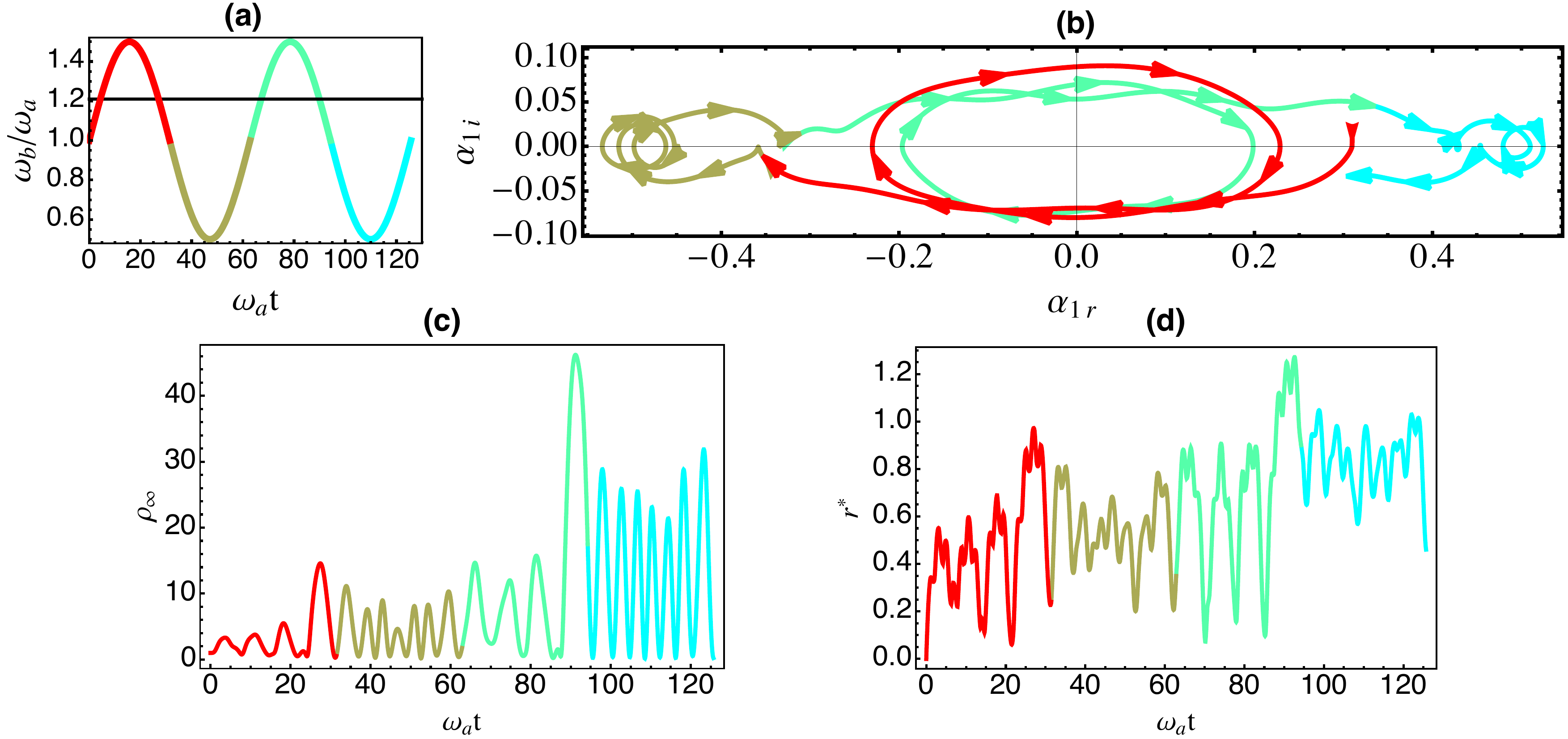}
\caption{ (Color online): We show the driving cycle [panel {\bf (a)}], the mean field trajectory {\bf (b)},
the photon number fluctuations $\rho_\infty$ {\bf (c)}, and the two-mode
squeezing parameter $r^*$ {\bf (d)}. In panels {\bf (b)}-{\bf (d)},
segments of a given color refer to the corresponding part of the driving cycle [panel {\bf (a)}]. We set $\eta = 0.1 \omega_a$ and
$\lambda=0.5$, with $\lambda_0=1$, and $g= 0.55 \omega_a$. Initial
conditions are $\alpha(0)=\alpha^{sr}(0)$ and $\beta(0)=\beta^{sr}(0)$.
        }
        \label{Fig mean field}
        \end{center}
\end{figure}

For a slow driving, the trajectories show some regularity as,
after an initial amplification stage, the mean fields remarkably
tend to be {\it attracted} towards the nearest of the equilibrium
points; namely, either $(\alpha^{n}, \beta^{n})$, or one of the
two broken symmetry points $(\alpha^{sr}_{\pm},
\beta^{sr}_{\pm})$, which is then `followed' as its value changes
due to the driving. More specifically, when $\mu (t)>1$, the two
mean fields circulate around their N stationary point,
$\alpha^{n}$ and $\beta^{n}$, respectively. A switching occurs
once $\mu(t)$ crosses its critical value, with the mean field
trajectory that dynamically breaks the parity symmetry, ``selects
a sign'', and moves towards either the positive or the negative
real axis, to begins rotating around one of SR values. When,
later, we get $\mu(t) >1$ again, the mean fields are attracted
back by the N fixed point, to move again towards one of the SR
points further on. Such a sequence of switching events occurs if
the driving parameters and the coupling strength are such that
$\mu_0 \in [(1+\lambda/\lambda_0)^{-1},1]$ whose boundaries
correspond to the black and red lines in Fig.~\ref{Fig1} {\bf
(a)}, respectively. This is the case, e.g., of Fig. ~\ref{Fig1}
{\bf (c)}. A different behavior is found for
$\mu_0 \in (1,(1-\lambda/\lambda_0)^{-1}]$, where the right
boundary corresponds to the green line in Fig. \ref{Fig1}~{\bf
(a)}. In this case, although $\mu<1$ for some $t$,  the trajectory
keeps rotating around the N equilibrium point, as in
Fig.~\ref{Fig1} {\bf (b)}. For parameters taking us below the
green line, the driving is not able to bring the system to
criticality and the solution of Eqs.~(\ref{eqs}) never exit the
linear regime. Correspondingly, the values of the mean fields
remain of order $\epsilon$. Above the black line, on the other
hand, only the two SR stationary points come into play. One
further particular trajectory for the mean photon field, in the
regime in which both the N and the SR points are relevant, is
analyzed in detail in Fig. (\ref{Fig mean field}), which shows how
the field follows the instantaneous equilibrium point, dynamically
breaking the equivalence between the two SR values when entering
the regime with $\mu < 1$.

The dynamic mean fields are not sufficient to obtain the average
value of a generic observable. This is the case, for instance, of
the photon number. A complete quantum description requires the
knowledge of the (operator) fluctuations around the mean fields.
The associated equations of motion for the fluctuations are
explicitly derived in~\cite{nota}. They are best displayed in
terms of the quadrature vector $\hat {\bm Q}(t)= (\hat q_c,\hat
q_d,\hat p_c,\hat p_d)$ with $\hat q_k= (\hat k+\hat
k^{\dag})/\sqrt 2$, $\hat p_k= i (\hat k^{\dag}-\hat k)/\sqrt 2$
($k=c,d$), and take the form
\begin{equation}
\dot{\hat{\bm Q}}(t)= {\bm M}_{\alpha, \, \beta} \,  \hat{\bm
Q}(t), \label{wveceq}
\end{equation}
where matrix ${\bm M}_{\alpha, \, \beta}$ is a non-linear function
of the instantaneous mean field values $\alpha$ and
$\beta$~\cite{nota}. Although the dynamics of the mean fields
discussed above is independent of the fluctuations, the reverse is
not true. Moreover, for $\alpha, \beta \ll 1$ (i.e. when the mean
fields are in the linear regime), ${\bm M}_{\alpha, \, \beta}$
reduces to the very same dynamical kernel ruling Eqs. (\ref{eqs})
for $t < t_{\text{lin}}$. This implies that the stability analysis
of Fig.~\ref{Fig1} applies to the quantum fluctuations as well. In
particular, for a parameter set in the white region, the dynamics
of the fluctuations becomes chaotic: two very close initial states
exponentially diverge in time, at a  rate $\gamma^*$.
Eq.~(\ref{wveceq}) are formally solved as $\hat{\bm Q}(t) =
\Phi(t) \hat{\bm Q}(0)$, where matrix $\Phi$ is such that
$\dot{\Phi} = {\bm M}_{\alpha, \, \beta} \Phi$ and satisfies the
boundary condition $\Phi(0) = \mathds{1}$. The first moments of
the quadratures are zero at all times (as the averages are given
by the mean fields), while the second moments form the covariance
matrix $W$ with elements $ W_{i\,j} = \langle\{\hat Q_i,\hat Q_j\}
\rangle/2 - \langle \hat Q_i \rangle \langle \hat Q_j\rangle$,
\cite{Weedbrook2012,Simon1994,Ferraro05arxiv}. The time-evolved
covariance matrix is then given by $W(t) =  \Phi(t) W(0) \Phi(t)^T
$. In general, both the mean fields and the quantum fluctuations
contribute to the evolution of a physical observable, an
interesting example being given by the average photon number
\begin{equation}\label{photons number}
n_a  =  \langle \hat a^{\dag} \hat a\rangle \equiv N
\abs{\alpha}^2+ \left( W_{11} + W_{33} -1\right)/2.
\end{equation}
Covariance matrix contributions dominate for initial mean fields
values $\lesssim N^{-1/2}$, and, in particular, if one takes the
vacuum of both modes as initial state. In fact, in virtue of the
previous discussion, this implies $\alpha=\beta=0 \, \forall t$,
so that photons are generated, in this case, by the exponential
amplification of the initial fluctuations, due to the instability
of the system under sinusoidal perturbation signalled by a
positive instability rate $\gamma^*$. For a very fast
perturbation, $\eta \gg \omega_a, \omega_b$, this requires
$\mu_0<1$. On the other hand, if $\mu_0 >1$, fluctuations are
bounded in time. Differently, for a very slow perturbation, photon
production occurs if we can make $\mu(t) <1$ in some time
interval, while fluctuations are bounded only if $\mu (t)
>1\,\forall t$.
Photon generation from the vacuum is related to the thermodynamic
work done on the system by the driving agent~\cite{work}. The
average work performed at time $t$ is
\begin{equation}
\begin{aligned}
\langle w \rangle &= 
\langle\hat H\rangle_t-\langle\hat H\rangle_0= \omega_a n_a(t)  + {N}\omega_b(0)/2+2gW_{12}(t)\\
&+ {\omega_b(t)}\left[ W_{22}(t) + W_{44}(t) -(N+1) \right]/2,
\end{aligned}
\label{dipew12}
\end{equation}
which depends not only on the local energies of the two modes
involved, but also on their correlations through the $W_{12}$
term. Eq. (\ref{dipew12}) shows that not all of the energy pumped
into the system is used for photon production in light of the
non-adiabatic nature of the driving. Part of such energy goes to
the atoms and part is stored as interaction energy. The
non-adiabaticity of the driving process can be quantitatively
studied using specifically designed thermodynamic figures of merit
for irreversibility~\cite{work}. Among them is the {\it inner
friction}~\cite{Plastina14PRL113}, which is defined as the
non-adiabatic part of the work. In our case, assuming the coupling
to be switched on at $t=0$, it reads $ \langle w_{fric} \rangle =
\langle w \rangle -  E^t_{GS} + E^0_{GS}$, where $E^t_{GS}$ is
instantaneous energy of the ground state $\ket{GS}$ of the system,
which becomes non-analytic for $\mu=1$. The behavior of $\langle
w_{fric} \rangle$ crucially depends on
the driving amplitude $\lambda$.
If $\lambda$ is such that $\mu_{min}>1$,
the system never exit the N region and $\langle w_{fric} \rangle$
never gets macroscopic values. If $\mu_{min}=1$
(i.e., $\mu=1$ at times $\eta t=(3/2 + 2 k )\pi$, $k\in
\mathbb{Z}^*$), the inner friction becomes non
analytic (as the energy gap closes), 
yet remaining ${\cal O}(1)$ as $N\rightarrow \infty$. If $\lambda$
is such that $\mu_{min} <1$,
instead, the system enters the SR region for $t \in [(2 k-1) \pi +
\tilde t, 2 k \pi -\tilde t]$, where $\eta \tilde t = \arcsin
(\lambda_c / \lambda) $ ($k\in\mathbb{Z}^*$). Then, as the evolved
state becomes macroscopically different from the instantaneous
ground state, inner friction gets macroscopic values. Altogether,
the inner friction per atom in the thermodynamic limit is
\begin{equation}
\lim_{N\rightarrow \infty}{\langle w_{fric} \rangle}/{N} =
\begin{cases}
0  & \text{for}~\mu(t) \geq 1, \\
{\omega_b(t)}[1-1/\mu(t)]^2 /4& \text{for}~\mu(t)<1.
\end{cases}
\end{equation}
Besides the mean photon number and the average (non-adiabatic)
work, we can characterize the photon statistics by the variance,
$\sigma_a^2 = \langle (a^\dag a)^2 \rangle - n_a^2$, and the
Mandel parameter $\rho = \frac{\sigma_a^2 }{n_a}$. The latter
signals a sub- or super-Poissonian statistics, with $\rho=1$ for a
coherent state. For very large $N$,  in the regime in which the
mean fields dominate with respect to quantum fluctuations (which
is the case for initial mean fields larger than $N^{-1/2}$),  and
neglecting $ {\cal O}(\sqrt{N})$ terms, the variance is
\begin{equation}\label{photons variance}
 \sigma_a^2 = 2 N \left( \alpha_{r}^2 W_{11} + \alpha_{i}^2 W_{33} + 2 \alpha_{r} \alpha_{i} W_{13} \right)
\end{equation}
so that, as $N \rightarrow \infty$, the Mandel parameter becomes
\begin{equation}\label{ratio poisson}
\rho_{\infty} = {2}(\alpha_{r}^2 W_{11} + \alpha_{i}^2 W_{33}
+ 2 \alpha_{r} \alpha_{i} W_{13})/|\alpha|^2
\end{equation}
In general terms, the time behavior of the quantum fluctuations is
very different depending on wether $\mu$ is larger or smaller than
$1$. This is reflected in the covariance matrix and witnessed by
the parameter $\rho_{\infty}$ (see Fig. \ref{Fig mean field}).
Roughly, this is an oscillating function of time, which oscillates
faster as the mean fields get larger, with an amplitude that
suddenly increases whenever $\mu$ crosses $\mu_c$ to enter in the
SR region. 


While $n_a$ and $\rho$ describe the reduced photon-state only, we
can also characterize global state correlations by evaluating the
degree of two-mode squeezing. With this aim, we consider the
parameter $r_{opt}(t)$ optimizing the fidelity
\cite{Marian12PRA86} between $\ket{\psi,t}$ and the two-mode
squeezed coherent state $\ket{\psi_{\alpha, \, \beta}(r,t)}$
having coherent amplitudes given by the mean fields and (real)
squeezing degree $r$, i.e.
$|{\psi_{\alpha, \beta}(r,t)}\rangle= e^{r ( c^\dag d^\dag - c d)
} \ket{\sqrt N \, \alpha(t)}_a\ket{\sqrt N \, \beta(t)}_b$.
Regardless of $t$, we find a value $r_{opt}(t)$ for which the
fidelity is $\ge 0.9999$~\cite{nota}. Therefore, $r_{opt}$ itself
can be thought as a good (although approximate) descriptor of the
photon-atom entanglement, \cite{Horodecki96PLA96,Simon00PRL84}.
During time evolution, it turns out that $r_{opt}$ can either grow
exponentially at short times (if $\gamma^*>0$), or remains very
close to its initial value (for $\gamma^*=0$). An example of the
behavior of $r_{opt}$ is reported in Fig. (\ref{Fig mean field}),
for the various stages of the driving induced dynamics. After the
initial fast increase, $r_{opt}$ displays small `jumps' whenever
the driving brings the system in the SR region.

\noindent {\it Concluding remarks.-}We have provided a dynamical
mean field-based description of the driven Dicke model, discussing
the non-linear evolution of the mean fields, as well as that of
quantum fluctuations. Both are needed to determine the time
behavior of physical observables such as the photon number. Our
approach is exact in the thermodynamic limit. However, for finite
$N$, our description is accurate provided that fluctuations do not
become macroscopic (otherwise the expansion in
Eq.~(\ref{expansion}) breaks down). This gives a time limit
$t_{max}$ (generically increasing with $N$) within which the
analysis is meaningful. Remarkably $t_{max}$, which is estimated
in~\cite{nota}, may be different in the various dynamical regimes.
Within such limit, we have discussed the phenomenon of dynamical
breaking of the parity symmetry in the mean field evolution under
driving, analysed photon generation from the vacuum, using
out-of-equilibrium thermodynamical tools to characterize it, and
described the generation of two-mode squeezing and entanglement
between field and atoms.

\noindent {\it Acknowledgements} We acknowledge support from the
Collaborative Projects QuProCS (Grant Agreement 641277), and
TherMiQ (Grant Agreement 618074), the John Templeton Foundation
(grant number 43467), the Julian Schwinger Foundation (grant
number JSF-14-7-0000), and the UK EPSRC (grant number
EP/M003019/1). We acknowledge partial support from COST Action
MP1209.

\section{Supplementary material}

\subsection{Holstein-Primakoff trasformation} Since the total spin
$|\vec J|^2$ is conserved, the full Hilbert space can be
decomposed into invariant subspaces labelled by the index $j$
describing its eigenvalues, $j=1,2,\cdots,\frac{N}{2}$ for an even
$N$, or $j=\frac{1}{2},\cdots,\frac{N}{2}$ if $N$ is odd. Defining
the projector $P_j$ onto $j$-th subspace, we can rewrite the
Hamiltonian as $H = \sum_j P_j H  P_j \equiv \sum_j H^{(j)} $

In the main text we focus on the case of a fully symmetric initial
atomic state, which implies selecting the $j=\frac{N}{2}$
invariant subspace. We therefore performed the Holstein-Primakoff
transformation on the spin operators projected in this sector
alone. The procedure, however, could have been repeated for each
$j$. To start with, one has to express the projected spin
$\vec{J}^{(j)}=P_{j} \vec{J} P_{j}$ through boson operators
$b_{j}, b_j^{\dag}$:
$$  J_z^{(j)} = b_j^\dag b_j - j  \, , \quad
  J_+^{(j)} = b_j^\dag \sqrt{2 j - b_j^\dag b_j}
$$
so that the projected Hamiltonian is
\begin{eqnarray}\label{hamiltonian HN}
  H^{(j)} &=& \omega_a a^\dag a + \omega_b \left( b_j^\dag b_j - j \right) + \\ \nonumber
   &&   g \left( a^\dag + a \right)\left(b_j^\dag\sqrt{1-2 \frac{b_j^\dag b_j}{j}}  + \sqrt{1-2\frac{b_j^\dag b_j}{j}}b_j \right)
\end{eqnarray}
In the main text, $j=N/2$ is taken and all of the subscripts are
erased.

\subsection{Equations of motion} The core of our
approximate analysis is the expansion of the non linear term
\begin{equation}\label{rootapp}
\sqrt{1-\frac{b^\dag b}{N}} = \sqrt{\Gamma}
\sqrt{1-\frac{\sqrt{N}(\beta d^\dag + \beta^* d) + d^\dag d}{N}}
\, .
\end{equation}
In the limit $N\rightarrow \infty$, this can be expanded into a
power series as
\begin{eqnarray}\label{expansionapp}
\sqrt{1-\frac{b^\dag b}{N}} &=& \sqrt{\Gamma} \bigg( 1 - \frac{\beta d^\dag + \beta^* d}{2 \sqrt{N} \Gamma} -\frac{d^\dag d}{2N \Gamma} \\
\nonumber && -\frac{\left(\beta d^\dag + \beta^* d\right)^2}{8 N
\Gamma^2} \bigg)+ O\left(N^{-\frac{3}{2}}\right)
\end{eqnarray}
with $\alpha(t)$ and $\beta(t)$ assumed to stay order $O(1)$ so
that the lowest order term gives already a good approximation once
inserted in the Hamiltonian.

Using the leading contribution only, the Hamiltonian becomes
\begin{eqnarray}
\nonumber H^{(N/2)} &=& \omega_a c^\dag c + \left[ \omega_b - 2 g
\frac{\alpha_{r} \beta_{r}}{\sqrt{\Gamma}}
\left(2+\frac{|\beta|^2}{2\Gamma}\right)
\right] d^{\dag} d + \\
&& g \sqrt{\Gamma}(c^\dag+c)\left[ \left(1-\frac{\beta^* \beta_{r}}{\Gamma}\right)d+h.c. \right]- \nonumber \\
&& g \frac{\alpha_{r}}{\sqrt{\Gamma}}\left[\beta^*\left(1+\frac{\beta^*\beta_{r}}{2 \Gamma}\right) d^2 + h.c.\right] + \nonumber \\
&& \sqrt{N}\left( \Delta_{c}^* c + \Delta_{d}^* d + h.c. \right) +
\Lambda_N + O\left(\frac{1}{\sqrt{N}}\right) \label{hamiltonian
evo}
\end{eqnarray}
where we defined $\Delta_{c}$ and $\Delta_{d}$ as
\begin{equation} \label{delta vec}
\vec{\Delta} = \left(
  \begin{array}{c}
    \Delta_{c} \\
    \Delta_{d} \\
  \end{array}
\right) = \left(
  \begin{array}{c}
    \omega_a \alpha + 2 g \sqrt{\Gamma} \beta_{r} \\
    \omega_b \beta + 2 g \sqrt{\Gamma} \alpha_{r}\left(1-\frac{\beta \beta_{r}}{\Gamma}\right) \\
  \end{array}
\right)
\end{equation}
while $\Lambda_N$ is the c-number
\begin{eqnarray}
\Lambda_N &=& N \left\{ \omega_a |\alpha|^2 + \omega_b \left(
|\beta|^2-
\frac{1}{2} \right )+4g \sqrt{\Gamma} \alpha_{r} \beta_{r} \right \}- \nonumber \\
&&  g\alpha_{r}\beta_{r} \frac{|\beta|^2}{2 \sqrt{\Gamma}}
\label{Lambda N}
\end{eqnarray}
In the strict thermodynamic limit, thus, the Hamiltonian becomes
quadratic, so that the time evolution can be described by a
Gaussian propagator.

We describe the dynamics in the Heisenberg picture, and obtain the
following equation for the annihilation operators:

\begin{eqnarray}
i \frac{d}{dt} c^{(H)}(t) &=& -
U^\dag(t,0)\left[H^{(N/2)}(t),c\right]U(t,0)
- i \sqrt N \frac{d \alpha(t)}{ dt} \label{heis eq} \\
\nonumber  &=&  \omega_a c^{(H)}  + g\sqrt{\Gamma}\left[\left(1-\frac{\beta^* \beta_{r}}{\Gamma}\right) d^{(H)} + h.c.\right] \\
\nonumber && + \sqrt{N} \Delta_{c}- i \sqrt N \frac{d \alpha}{ dt}
+ O\left(\frac{1}{\sqrt{N}}\right)
 \end{eqnarray}
In order for the operator $c$ to stay of order $O(1)$, the two
terms $\sim \sqrt N$ above should compensate each other. To this
end, we require the mean field $\alpha$ to satisfy

 \begin{equation}\label{eq diff alfa}
 i \frac{d\alpha}{dt} = \Delta_{c}
 \end{equation}

In this way, the Heisenberg equation becomes

 \begin{equation}\label{eq Heis c}
 i \frac{d}{dt}c^{(H)} =  \omega_a c^{(H)}  + g\sqrt{\Gamma}\left[\left(1-\frac{\beta^* \beta_{r}}{\Gamma}\right) d^{(H)} + h.c.\right] + O\left(\frac{1}{\sqrt{N}}\right)
 \end{equation}

In the limit $N \rightarrow \infty$ the terms
$O\left(\frac{1}{\sqrt{N}}\right)$ do not bring any contribution
to the dynamics of the operator $c^{(H)}$. For finite $N$, on the
other hand, in order to give a more accurate description of the
dynamics of the fluctuations, one should consider further terms in
the expansion \eqref{expansionapp}. Even if this is done, however,
the dynamics of the mean field will remain unchanged, as it is
determined by the $\sim \sqrt N$ terms only.

After a similar analysis is carried out for the operator $d$, we
find the differential equation for $\beta$

\begin{equation}\label{eq diff beta}
i \frac{d\beta}{dt} = \Delta_{d}
\end{equation}

and the Heisenberg equations

 \begin{eqnarray}
 \nonumber i \frac{d}{dt}d^{(H)} &=& \left( \omega_b - 2 g \frac{\alpha_{r}\beta_{r}}{\sqrt{\Gamma}}\left(2+\frac{|\beta|^2}{2 \Gamma}\right)\right) d^{(H)}\\
 \nonumber && +  g \sqrt{\Gamma}\left( 1 - \frac{\beta \beta_{r}}{\Gamma} \right)(c^{(H)} + h.c.)  \\
 \label{eq Heis d} && - 2g \frac{\alpha_{r}\beta}{\sqrt{\Gamma}}\left( 1 + \frac{\beta \beta_{r}}{2 \Gamma} \right) {d^{(H)}}^\dag + O\left(\frac{1}{\sqrt{N}}\right)
 \end{eqnarray}

\subsection{Time evolution of the mean fields} Explicitly, the
equations for the real and imaginary parts of the re-scaled mean
fields are

\begin{equation}
 \begin{cases}
        \dot{\alpha}_{r} = \omega_a \alpha_{i} \\
        \dot{\alpha}_{i} =  -\omega_a \alpha_{r} -2 g \sqrt{\Gamma} \beta_{r} \\
        \dot{\beta}_{r} = \omega_b \beta_{i} - 2g \alpha_{r} \frac{\beta_{r} \beta_{i}}{\sqrt{\Gamma}} \\
        \dot{\beta}_{i} = -\omega_b \beta_{r} - 2g\sqrt{\Gamma} \alpha_{r}(1-\frac{\beta_{r}^2}{\Gamma})
      \end{cases}
      \label{system}
\end{equation}
If we regard $\alpha_{r}$ and $\beta_{r}$ as the generalized
coordinates, and  $\alpha_{i}$ and $\beta_{i}$  as their conjugate
momenta, these equations can be derived from the Hamiltonian
function

\begin{equation*}
  H(\alpha,\beta) = \frac{\omega_a }{2}|\alpha|^2 + \frac{\omega_b }{2}|\beta|^2 + 2 g \sqrt{\Gamma}\beta_{r}\alpha_{r}
\end{equation*}
They can be considered as purely classical equations, but should
be solved with initial conditions that comes from the choice of an
initial quantum state:

 \begin{equation}\label{initial conditions}
 \left(
                  \begin{array}{c}
                    \alpha(0) \\
                    \beta(0) \\
                  \end{array}
                \right) = \frac{1}{\sqrt{N}}\left(
                  \begin{array}{c}
                    \langle a(0) \rangle \\
                    \langle b(0) \rangle \\
                  \end{array}
                \right)
 \end{equation}

As our treatment is based on the requirement that fluctuations
stay of lower order than averages, we need to restrict our-selves
to initial quantum states that fulfill this very same condition.

The system \eqref{system} can be linearized if the initial
conditions are such that $|\alpha(0)|,|\beta(0)| \ll 1$. Then,
until the time $t_{lin}$ such that $\alpha$ and $\beta$ are of
order one, the dynamics of the mean fields can be described by the
linearized system

\begin{equation}\label{system diff linear}
\frac{d}{dt}\left(
  \begin{array}{c}
    \alpha_{r} \\
    \beta_{r} \\
    \alpha_{i} \\
    \beta_{i} \\
  \end{array}
\right) \approx M_{0}\left(
  \begin{array}{c}
    \alpha_{r} \\
    \beta_{r} \\
    \alpha_{i} \\
    \beta_{i} \\
  \end{array}
\right)
\end{equation}

where

\begin{equation}\nonumber
M_{0}=\left(
            \begin{array}{cccc}
              0 & 0 & \omega_a & 0 \\
              0 & 0 & 0 & \omega_b \\
              -\omega_a & -2g & 0 & 0 \\
              -2g & -\omega_b & 0 & 0 \\
            \end{array}
          \right)
\end{equation}

\subsection{Time evolution of the fluctuations} As for the quantum
fluctuations, we can recast the Heisenberg equations \eqref{eq
Heis c} and \eqref{eq Heis d} in a more compact form using the
quadrature operators $\vec Q = (q_c,q_d,p_c,p_d)$ as introduced in
the main text. In the limit $N\rightarrow \infty$, we find

\begin{equation}\nonumber
\dot{\vec{Q}}^{(H)} = M_{\alpha,\beta} \vec{Q}^{(H)}
\end{equation}

where $M_{\alpha,\beta}$ is the matrix

\begin{widetext}
\begin{equation} \label{matrix M}
M_{\alpha,\beta} = \left(
      \begin{array}{cccc}
        0 & 0 & \omega_a & 0 \\
-\frac{2 g \beta_{r}\beta_{i} }{\sqrt{\Gamma}} & - \frac{2g
\alpha_{r} \beta_{i}}{\sqrt{\Gamma}}\left( 1 +
\frac{\beta_{r}^2}{\Gamma}\right) &
0 & \omega_b - \frac{2 g \alpha_{r} \beta_{r}}{\sqrt{\Gamma}}\left( 1+ \frac{\beta_{i}^2}{\Gamma} \right) \\
-\omega_a & -2g\sqrt{\Gamma}\left(1-\frac{\beta_{r}^2}{\Gamma}\right) & 0 & 2g\frac{\beta_{r}\beta_{i}}{\sqrt{\Gamma}} \\
-2g\sqrt{\Gamma}\left(1-\frac{\beta_{r}^2}{\Gamma}\right) &
-\omega_b + \frac{2g\alpha_{r}\beta_{r}}{\sqrt{\Gamma}}\left( 3 +
\frac{\beta_{r}^2}{\Gamma} \right) &
0 & \frac{2 g \alpha_{r} \beta_{i}}{\sqrt{\Gamma}}\left( 1+\frac{\beta_{r}^2}{\Gamma} \right) \\
      \end{array}
    \right)
\end{equation}
\end{widetext}

We can define the fundamental matrix $\Phi$ as the solution of the
differential equation
\begin{equation} \label{phi super}
\dot{\Phi} = M_{\alpha,\beta} \Phi
\end{equation}
with the initial condition $\Phi(0) = \mathds{1}$. Then the
solution of the Heisenberg equation is
\begin{equation} \nonumber
\vec{Q}^{(H)}(t) = \Phi(t) \vec{Q}
\end{equation}

In our description the first moments are zero for every time, i.e.
$\langle \vec{Q} \rangle = \vec{0}$. The second moments are given
by the covariance matrix $W$, as mentioned in the main text. If
the initial state is Gaussian, then the covariance matrix
completely characterizes the fluctuations around the mean fields.

In the linear transient regime, $t < t_{lin}$, we find that
$M_{\alpha,\beta} = M_0 $, so that the dynamics of the
fluctuations are the same as the mean field ones. Furthermore, if
the initial state is such that $\alpha(0)=\beta(0)=0$, then the
mean fields will remain zero $\forall t$ and, thus the dynamics of
the fluctuations is described by $M_0$ at all times.

In general, $M_{\alpha, \beta}$ has a parametric time dependence,
as it contains $\omega_b(t)$. Therefore, it inherits from
$\omega_b$ a periodicity of $T=2 \pi/\eta$. Then, by the Floquet
theorem, we can write
$$\Phi(t) = e^{-\frac{B t }{T}} \, P(t) \, , $$
with a constant $B$ and a periodic $P(t)$. For us, $\Phi
(0)=\mathds{1}$, therefore also $P(0) = P(T) = \mathds{1}$. As a
result, we have that the so called monodromy matrix is $\Phi(T) =
e^{- B}$.

\subsection{Limits of validity for finite $N$}

Our description  of the dynamics becomes exact for every finite
time $t$ in the limit $N\rightarrow \infty$. Anyway, it's crucial
to understand what the limits of applicability of our approach are
for finite $N$. We observe that, during the time evolution,
fluctuations (assumed to be  $O(1)$ at $t=0$) can become very
large and even unbounded. When this happens, the expansion of the
square root in the Hamiltonian may become incorrect. In other
words, for finite but large $N$ our description stays accurate
until the n-th moments become of order $N^{n/2}$. To ensure that
this is not the case, it is enough to require that all of the
elements of the matrix $\Phi$ are small compared to $\sqrt{N}$.
Therefore, our description of the dynamics is accurate until a
time $t_{max}$, defined as the first instant for which

\begin{equation}\label{condition time}
max_{i\,j} \left\{ \Phi_{ij}(t_{max}) \right\} \approx \sqrt{N}
\end{equation}

In the limit $N\rightarrow \infty$, we expect $t_{max} \rightarrow
\infty$.

In order to characterize and estimate $t_{max}$, we need to
consider the initial conditions for the mean fields, $\alpha(0) =
\alpha_0$ and $\beta(0) = \beta_0$. In particular, if we take
$\delta= max\left\{|\alpha_0| , |\beta_0|\right\} \ll 1$, the
first part of the dynamics is included in the linear transient.
This implies that, for $0\leq t <t_{lin}$, the time evolution of
both the fluctuations and of the mean-fields is determined by the
linearized matrix $M_0$.

In the absence of driving and in the linear transient, the
dynamics would be characterized by the eigenvalues of the matrix
$M_0$. Two of them are always purely imaginary, i.e. $\pm i
\lambda_1$. The other two are $\pm i \lambda_2$, where $\lambda_2$
is real if $\mu>1$, it is equal to $\lambda_1$ if $\mu=1$ and it
is purely imaginary if $\mu<1$. This means that if $\mu>1$ the
fluctuations and the mean fields stay always of the same order,
i.e $t_{lin}\rightarrow \infty$ and $t_{max}\rightarrow \infty$.
At the transition point, $\mu=1$, fluctuations and mean-fields
grow linearly in time; while in the super-radiant phase $\mu<1$,
the fluctuations and the mean fields can experience an exponential
growth, with an {\it instability rate} given by
$\gamma^*=|\lambda_2|$. This is true until $t\lesssim t_{lin}$.
After $t_{lin}$, the nonlinear terms cannot be neglected anymore
and one has to consider the full non-linear equations. Thus,
within the linear regime, we can define the characteristic time
$\tau^*$ as the inverse of the instability rate, i.e.

\begin{equation}\nonumber
{\tau^*}^{-1} = \gamma^*=
\sqrt{\sqrt{\left(\frac{\omega_a^2-\omega_b^2}{2}\right)^2+4\omega_a\omega_b
g^2}-\frac{\omega_a^2+\omega_b^2}{2}}
\end{equation}
We can estimate $t_{lin}$ as the time for which one of the mean
fields (either $\alpha$ or $\beta$) becomes of order one,
\begin{equation}\label{time Tstar}
t_{lin} \approx \tau^* \ln\left(\delta^{-1}\right)
\end{equation}
Since $\delta \ll 1$, we expect $t_{lin}$ to be much larger than
$\tau^*$. The crucial point, however, is wether or not all of the
linear transient is contained within our limit of validity.
Indeed, for this description to be valid for times of the order of
$t_{lin}$, we have to require that

\begin{equation}\nonumber
 e^{\gamma^* t_{lin}} \ll \sqrt{N}
\end{equation}

from which it follows that

\begin{equation}\nonumber
 t_{lin} \ll \frac{\tau^*}{2} \ln(N)
\end{equation}

and that

\begin{equation}\nonumber
\delta \gg \frac{1}{\sqrt{N}}
\end{equation}

This means that if the initial state is not so close to the vacuum
(the difference from zero of $\alpha$ and $\beta$ being larger
than $\frac{1}{\sqrt{N}}$), then $t_{lin} \ll t_{max}$. In this
case, our description makes full sense even outside the linear
transient, and can be used even when the mean-fields take
macroscopic values. In this case, $t_{lin}$ can indeed be
estimated by \eqref{time Tstar} and is finite. On the other hand,
for finite $N$, it is not possible to give a simple expression for
$t_{max}$, as it depends on the non-linear terms appearing in the
time evolution, and the only way to check the validity of our
approach is to check the condition \eqref{condition time}.

Instead, if both of the mean fields start too close to zero,
specifically if $\delta \lesssim \frac{1}{\sqrt{N}}$, then our
description ceases to be good before the mean fields $\alpha$ and
$\beta$ reach macroscopic values. In this case, in fact, $t_{max}$
is inside the linear transient, and, from \eqref{condition time},
we can make the simple estimate

\begin{equation}\nonumber
t_{max} = \frac{\tau^*}{2}\ln(N)
\end{equation}

These estimates and reasoning can be adapted also to the case of a
driven system, since they are simply a consequence of the fact
that the dynamics of the mean fields and that of the fluctuations
are the same in the linear transient.

For instance, for a periodically driven Hamiltonian, we can use
the largest positive Floquet exponent of the linearized matrix
$M_0$ in order to estimate $\tau^*$; and then use again the
equations above to estimate $t_{lin}$ and $t_{max}$. Specifically,
if $M_0$ is periodic in time, with period $T$, then the monodromy
matrix (in the Floquet description) is given by
$\mathcal{M}=\Phi(T)$. The eigenvalues of $\mathcal{M}$ are the
Floquet multipliers $\{\rho_i\}_{i=1}^4$. From these, we can
calculate the Floquet exponents, that are the complex numbers
$\nu_i = \frac{\ln(\rho_i)}{T}$. So we can define the instability
rate $\gamma^*$ as the maximum among zero and the real parts of
the Floquet exponents, i.e.

\begin{equation}\label{time tau star}
   \gamma^* = max \left\{ 0,\, \left\{\mathds{R}e\{\nu_i \} \right\}_{i=1}^4\right\}
\end{equation}

As a result, the arguments above can be applied in this case too,
even if, in a strict sense, the stability of the dynamics cannot
be fully characterized by the instantaneous eigenvalues of the
matrix $M_0$.


\subsection{Optimal degree of two-mode squeezing}
\begin{figure}[h]
        \begin{center}
        \includegraphics[width=\linewidth]{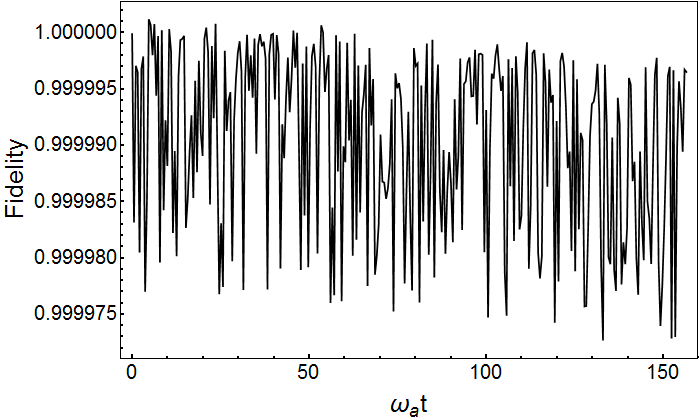}
        \caption{
The fidelity $F(t) = |\left \langle
\psi_{\alpha(t),\beta(t)}(r_{opt},t) | \Psi, t \right \rangle
|^2$, evaluated for the same parameter values used in Fig. 2 of
the main text. $F(t)$ stays very close to one, tending to slowly
decrease for long times.}
        \label{Fig fidelity}
        \end{center}
\end{figure}

As discussed in the main text, we compare the instantaneous state
of the global (atom+field) system with the two-mode squeezed
coherent state $\ket{\psi_{\alpha,\beta}(r,t)}$ obtained by
applying the unitary squeeze operator of (real) degree $r$,
$$S(r)= e^{r ( c^\dag d^\dag - c d) } \, ,$$
to the two-mode coherent state obtained by taking the
instantaneous mean fields $\alpha(t)$ and $\beta(t)$ as
amplitudes,
\begin{equation}
\ket{\psi_{\alpha,\beta}(r,t)}=S(r)\ket{\sqrt N \,
\alpha(t)}_a\ket{\sqrt N \, \beta(t)}_b
\end{equation}
The fidelity that we obtain by optimizing the parameter $r$, is
very close to unity, as reported in Fig. (\ref{Fig fidelity}).

\begin{thebibliography}{12}


\bibitem{polkormp}
A. Polkovnikov, K. Sengupta, A. Silva, and M. Vengalattore, Rev.
Mod. Phys. {\bf 83}, 863 (2011).

\bibitem{eisertnatphys}
J. Eisert, M. Friesdorf, and C. Gogolin, Nat. Phys. {\bf 11}, 124
(2015).

\bibitem{varikz}
B. Damski, Phys. Rev. Lett. {\bf 95}, 035701 (2005) ; W. H. Zurek,
U. Dorner, and P. Zoller, Phys. Rev. Lett. {\bf 95}, 105701
(2005); J. Dziarmaga, Phys. Rev. Lett. {\bf 95} , 245701 (2005);
B. Damski and W. H. Zurek, New J. Phys. {\bf 10} , 045023 (2008);
J. Dziarmaga, Adv. Phys. {\bf 59}, 1063 (2010); C. De Grandi, V.
Gritsev, and A. Polkovnikov, Phys. Rev. B {\bf 81}, 012303 (2010);
P. Silvi, G. Morigi, T. Calarco, S. Montangero Phys. Rev. Lett.
{\bf 116}, 225701 (2016).

\bibitem{Brennecke} K. Baumann, C. Guerlin, F. Brennecke, and T. Esslinger,
Nature (London) {\bf 464}, 1301 (2010).

\bibitem{pnas} J. Klinder, H. Ke\ss ler, M. Wolke, L. Mathey, A.
Hemmerich, Proc. Nat. Ac. Sc. {\bf 112}, 3290 (2015).

\bibitem{dicke} R. H. Dicke, Phys. Rev. {\bf 93}, 99 (1954); K. Hepp
and E. H. Lieb, Ann. Phys. (N.Y.) {\bf 76}, 360 (1973); Y. Wang
and F. Hioe, Phys. Rev. A {\bf 7}, 831 (1973); C. Emary and T.
Brandes, Phys. Rev. Lett. {\bf 90}, 044101 (2003); T. Brandes,
Phys. Rep. {\bf 408}, 315 (2005);  J. Vidal and S. Dusuel,
Europhys. Lett. {\bf 74}, 817 (2006); F. Plastina, G. Liberti, and
A. Carollo, Europhys. Lett. {\bf 76}, 182 (2006); G. Liberti, F.
Plastina, and F. Piperno, Phys. Rev. A {\bf 74}, 022324 (2006); Q.
H. Chen, Y. Y. Zhang, T. Liu, and K. L. Wang, Phys. Rev. A {\bf
78}, 051801(R) (2008); G. Liberti, F. Piperno, F. Plastina, Phys.
Rev. A {\bf 81}, 013818 (2010).

\bibitem{varie} J. Gong, L. Morales-Molina, and P. H\"anggi, Phys. Rev.
Lett. {\bf 103}, 133002 (2009); A. Eckardt, C. Weiss, and M. Holthaus, Phys. Rev. Lett.
{\bf 95}, 260404 (2005).  H. Lignier, C. Sias, D. Ciampini, Y. Singh, A. Zenesini,
O. Morsch, and E. Arimondo, Phys. Rev. Lett. {\bf 99}, 220403
(2007); G. G\"unter {\it et al.}, Nature (London) {\bf 458}, 178 (2009).


\bibitem{Bastidas12PRL108}
V. M. Bastidas, C. Emary, B. Regler, and T. Brandes, Phys. Rev.
Lett {\bf 108}, 043003 (2012).

\bibitem{plenion=1}
M.-J. Hwang, R. Puebla, and M. B. Plenio, Phys. Rev.
Lett. {\bf 115}, 180404 (2015).


\bibitem{Emary03PRE67}
C. Emary, and T. Brandes, Phys. Rev. E {\bf 67}, 066203,
(2003).


\bibitem{nota}
Supplementary Material.

\bibitem{Vacanti12PRL108}
G. Vacanti,  S. Pugnetti, N. Didier, M. Paternostro, G. M. Palma,
R. Fazio,V. Vedral, Phys. Rev. Lett. {\bf 108},
093603 (2012).

\bibitem{Floq} J. H. Shirley, Phys. Rev. {\bf 138}, B979 (1965).









\bibitem{Husimi53PTP9}
K. Husimi, {\it Miscellanea in Elementary Quantum Mechanics II},
Prog. Theor. Phys. {\bf 9}, 381, (1953).

\bibitem{Weedbrook2012} C. Weedbrook, and S. Pirandola, Rev. Mod. Phys. {\bf 84}, 621
(2012).

\bibitem{Simon1994} R. Simon, N. Mukunda, and B. Dutta, Phys. Rev. A,  {\bf 49}, 1567 (1994).

\bibitem{Ferraro05arxiv} A. Ferraro, S.
Olivares, and M. G. A. Paris, {\it Gaussian states in continuos variable quantum information}, Bibliopolis (Napoli, 2005).
(2005).

\bibitem{work}
M. Campisi, P. H\"{a}nggi, and P. Talkner, Rev. Mod.
Phys. {\bf 83}, 771 (2011).

\bibitem{Plastina14PRL113} R. Kosloff and T. Feldmann, Phys. Rev. E {\bf 61}, 4774 (2000); F. Plastina, A. Alecce, T. J. G.
Apollaro, G. Falcone, G. Francica, F. Galve, N. Lo Gullo, and R.
Zambrini, Phys. Rev. Lett. {\bf 113}, 260601
(2014).

\bibitem{Marian12PRA86}
P. Marian and T. A. Marian, Phys. Rev. A, {\bf 86}, 022340 (2012).

\bibitem{Horodecki96PLA96} M. Horodecki, P. Horodecki, and R.
Horodecki, Phys. Lett. A, {\bf 96}, 9601 (1996).

\bibitem{Simon00PRL84}
  R. Simon,
  Phys. Rev. Lett. {\bf 84}, 2726 (2000)


%
%




\end{thebibliography}
\end{document}